\documentclass[runningheads]{llncs}
\usepackage{graphicx}
\usepackage[linesnumbered,ruled,scleft,nofillcomment,vlined]{algorithm2e}
\usepackage{multirow}
\usepackage{adjustbox}
\usepackage{subcaption}
\begin{document}
\title{Space- and Time-Efficient Storage of LiDAR Point Clouds\thanks{This research has received funding from the European Union’s Horizon 2020 research and innovation programme under the Marie Sklodowska-Curie [grant agreement No 690941]; from the Ministerio de Econom{\'i}a y Competitividad (PGE and ERDF) [grant numbers TIN2016-78011-C4-1-R; TIN2016-77158-C4-3-R]; and from
Xunta de Galicia (co-founded with ERDF) [grant numbers ED431C 2017/58; ED431G/01].}}
%
%\titlerunning{Abbreviated paper title}
% If the paper title is too long for the running head, you can set
% an abbreviated paper title here
%
\author{Susana Ladra\inst{1}\orcidID{0000-0003-4616-0774} \and
Miguel R. Luaces\inst{1}\orcidID{0000-0003-0549-2000} \and
Jos{\'e} R. Param{\'a}\inst{1}\orcidID{0000-0002-8727-0980}\and
Fernando Silva-Coira\inst{1}\orcidID{0000-0003-1341-3368}}
\authorrunning{S. Ladra et al.}
% First names are abbreviated in the running head.
% If there are more than two authors, 'et al.' is used.
%
\institute{Universidade da Coruña, Fac. Inform\'atica, CITIC, Spain\\
\email{\{susana.ladra, luaces, jose.parama, fernando.silva\}@udc.es}}
\maketitle              % typeset the header of the contribution
\begin{abstract}
LiDAR devices obtain a 3D representation of a space. Due to the large size of the resulting datasets, there already exist storage methods that use compression and present some properties that resemble those of compact data structures. Specifically, LAZ format allows accesses to a given datum or portion of the data without having to decompress the whole dataset and provides indexation of the stored data. However, LAZ format still have some drawbacks that should be faced. 

In this work, we propose a new compact data structure for the representation of a cloud of LiDAR points that supports efficient queries, providing indexing capabilities that are superior to those of LAZ format.

\end{abstract}
\section{Introduction}\label{sec:introduction}

Light Detection and Ranging Technology (LiDAR) has been used during the last four decades in Geosciences as a geomatic method to obtain the 3D geometry of the surface of objects~\cite{dong2017lidar}. LiDAR uses a laser beam to compute the distance between a device and an object that reflects the beam. When the laser is used to scan the entire field of view at a high speed, the result is a dense cloud of points centered at the device, with each point having additional information such as the intensity of the laser beam reflection. If the device has additional sensors, each point in the cloud will have additional attributes (e.g., if the device includes a camera, each point will have a color value associated).

The decrease in cost of laser scanning devices has helped to drastically increase the application fields for point clouds. For instance, laser scanning has been used to classify and recognize objects in urban environments~\cite{wang2018lidar}, in natural environments (e.g., landslides~\cite{jaboyedoff2012use}, or forests~\cite{hyyppa2018forest}), or even underwater environments~\cite{palomer20183d}. Hence, huge datasets are being produced that require immense computing resources to be processed. To give two examples, the ISPRS benchmark on indoor modelling consists of five point clouds containing $98.1 \times 10^{6}$ points~\cite{khoshelham2017isprs} and the mobile laser scanning test data ``MLS 1 - TUM City Campus'' contains more than $1.7 \times 10^{9}$ points collected in 15 minutes.

Due to the size of the obtained data, the use of compression is almost mandatory. However, the traditional approach of keeping the data compressed in disk and decompressing the whole dataset before any processing is not effective. Therefore, the classical method for storing LiDAR data (LAZ format) is able to compress the data but, in addition, permits accessing a datum or portions of the data without the need of decompressing the whole dataset. In addition, it is equipped with an index to accelerate the queries. Observe that this setup is very similar to that of many modern compact data structures  \cite{Navarrobook} . 

The use of compression for spatial data is not exclusive of LAZ format. In the case of raster data, Geo-Tiff and NetCDF are able to store the data in compressed form, and   in the case of NetCDF, it also permits querying the data directly in that format. Therefore, the application  of the knowledge acquired in compact data structures soon led to a new research line.

The wavelet tree \cite{Grossi:2003} was the first compact data structure used in the scope of spatial data. The work in \cite{BLNSsecogis09} proposes a new point access method based on it. In  \cite{NNRtcs13}, it is used to represent a set of points in the two-dimensional space, each one with an associated value given by an integer function. Another family of compact data structures based on the quadtree also arose. In \cite{BernardoRoca13}, a compact version of the region quadtree was adapted for storing and indexing compressed rasters. In the same line, Ladra et. al \cite{LadraPS17} presented an improvement on the previous works. 

In this work we continue in this line, now tackling even more complex spatial data. We present a compact data structure to represent LiDAR point clouds, denoted $k^3$-$lidar$, which compresses and indexes the data. The improvements with respect to the LAZ format are in two aspects. The LAZ format relies on differential encoding plus an entropy encoder, which compresses/decompresses data by blocks, and thus, in order to retrieve a small region, one or more complete blocks must be decompressed. Another drawback of the LAZ format is that it uses a quadtree to index the points, and this only accelerates the queries by the $x$ and $y$ coordinates. Our new $k^3$-lidar is able to retrieve/decompress a given datum and indexes the three dimensions of the space. 

\section{Related work: LAS and LAZ format}\label{sec:related-work}
%\subsection{}

The American Society for Photogrammetry and Remote Sensing\footnote{https://www.asprs.org/} defined in 2003 the LASer (LAS) file format, an open data exchange format for LiDAR point data records~\cite{las_10}. The format contains binary data consisting of a header block that describes general information of the point cloud (e.g., number of points, bounding box), variable-length records to describe additional information such as georeferencing information or other metatada, and point records. Each point record consists of a collection of fields describing the point (e.g., the point $x$, $y$ and $z$ coordinates represented as 4-byte integers, the return intensity represented as a 2-byte integer, or the laser pulse return number). The standard defines a Point Data Record Format 0, and allows for additional formats to be defined with additional data (e.g., the Point Data Record Format 1 adds the GPS time at which the point was acquired as an 8-byte double).

The version 1.4 R14 of the LAS file format specification has been released in 2019~\cite{las_14r14}. It defines 11 point data record formats (some of them to provide legacy support), a mechanism to customize the LAS file format to meet application-specific needs by adding point classes and attributes, extended variable-length records to carry larger payloads, and additional types of variable-length records (e.g., georeferencing using Well Known Text descriptions, textual description of the LAS file content, or extra bytes for each point record).

Even though the LAS file format avoids using unnecessary space by storing the coordinates as scaled and offset integers, a LAS file requires much storage space (e.g., a 13.2M point cloud requires 254 MB of disk space, see Table~\ref{tab:experimental:datasets}). The LAZ file format~\cite{isenburg2013laszip}, defined by the LASzip lossless compressor for LiDAR, achieves high compression rates supporting streaming, and random access decompression. LASzip encodes the points in the cloud using chunks of 50,000 points. For each chunk, the first point is stored as raw bytes and it is used as the initial value for subsequent prediction schemes. Each additional point is compressed using an entropy coder (an adaptive, context-based arithmetic coding~\cite{sayood2002lossless}). These techniques make the LAZ file format very efficient in terms of space.  Isenburg~\cite{isenburg2013laszip} showed that the compression ratio is similar or better than general-purpose compression formats such as ZIP or RAR, while maintaining the possibility of processing the file as a stream of points or directly accessing a particular point without having to decompress the complete file.

The LAZ file format is also very efficient answering range queries on the $x$ and $y$ dimensions because LAZ files can be indexed using an adaptative quadtree over these coordinates. Each quadtree leaf contains a list of point indexes that can be used to determine the chunk that contains the point. To resolve a range query, the quadtree is first traversed to determine the candidate point indexes, then, the relevant chunks are retrieved, and finally the chunks have to be decompressed and sequentially scanned to determine the points in the result. 

Considering additional types of queries, the LAZ file format is highly inefficient on three-dimensional queries or queries over attribute data because a sequential scan has to be performed over the points in a chunk. These queries are becoming more common because classification algorithms over point clouds quite often require to locate close points in the three-dimensional space, or close points in a two-dimensional space with a similar attribute value (e.g., having the same intensity). Both types of queries can be efficiently answered if a three-dimensional index is built over the data.

%\subsection{Octree}

\section{Background: $k^2$-trees and $k^3$-trees}

The $k^2$-tree \cite{Brisaboa14} is a time- and space-efficient version of a region quadtree \cite{Klinger1971303,Samet:1984}. Considering a binary matrix of size $n \times n$, it is divided into $2^2$ \textit{quadboxes} (submatrices) of size $n/2 \times n/2$. Each quadbox produces a child of the root node. The label of the node is 1-bit if the corresponding quadbox contains at least one 1-bit, and 0-bit otherwise. The quadboxes having at least one 1-bit are divided using the same procedure until reaching a quadbox full of 0-bits, or reaching the cells of the original matrix.

The $k^2$-tree, instead of using a classical pointer-based representation of the tree,  represents the quadtree using only sequences of bits. More concretely, it uses two bitmaps, denoted as $T$ and $L$, where $T$ is formed by a breadth-first traversal of the internal nodes, whereas the $L$ is formed by the leaves of the tree. From this basic version, several other improvements yield better space and time performance \cite{Brisaboa14}. Among them, one is that instead of diving each quadbox into $2^2$  quadboxes of size $n/2 \times n/2$, each division produces $k^2$ quadboxes of size $n/k \times n/k$, where $k$ is a parameter that can be adapted for each level of the tree. This is usually used to obtain shorter and wider trees that, at the price of a slightly worse space consumption, are faster when querying.

\begin{figure}[th]
  \begin{center}
  \begin{tabular}{c}	
  \includegraphics[width=1\textwidth]{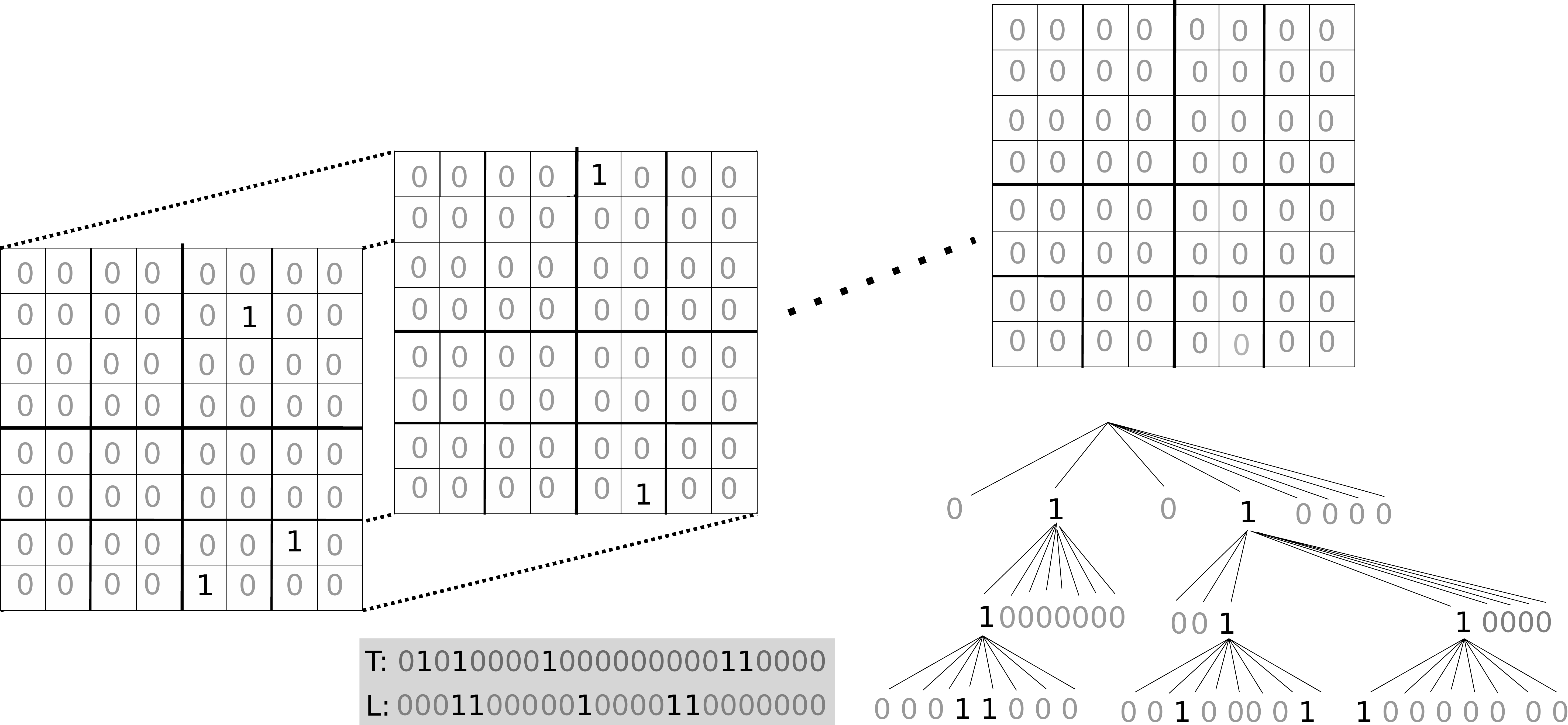}\\
  \end{tabular}
  \end{center}
  \vspace{-5mm}
  \caption{$k^3$-tree.}
\label{fig:k3tree}
\end{figure}

As in the case of the quadtree, where the simple addition of a third dimension produces the octree \cite{MEAGHER1982129}, the $k^3$-tree \cite{BernardoRoca13} is  simply a 3-dimensional $k^2$-tree. Figure \ref{fig:k3tree} shows a 3-dimensional binary matrix, its corresponding octree, and the $k^3$-tree represented using bitmaps $T$ and $L$.

The $k^3$-tree can be efficiently navigated using \textit{rank} and \textit{select} operations\footnote{Given a bitmap $B$, $rank_b(B,i)$ is the number of occurrences of bit $b$ in $B[1,i]$ and $select_b(B,j)$ is the $j$-th occurrence of bit $b$ in $B$.} over $T$ and $L$ (see ~\cite[Section 6.2.1]{BernardoPhD}).

%%%%%%%%%%%%%%%%%%%%%%%%%%%%%%%%%%%%%%%%%%%%%%%%%%%%%%%%%%%%%%%%%%%%%%
\section{Our proposal: $k^3$-lidar}\label{sec:k3-lidar}
%%%%%%%%%%%%%%%%%%%%%%%%%%%%%%%%%%%%%%%%%%%%%%%%%%%%%%%%%%%%%%%%%%%%%%

In this section, we present the $k^3$-lidar, a data structure to represent LiDAR point clouds in compact space, which  allows us to perform  efficient queries.

\subsection{Conceptual description}\label{sec:k3-lidar:conceptual}

Consider a 3D matrix of size $n \times n \times n$ that stores a set of LiDAR points. In addition to its coordinates, a point contains several values that correspond to each of its attributes (e.g., intensity, scan angle, color, etc.). 

Observe that the $k^3$-tree was designed to store and index a matrix with a bit value for all positions, that is, a 3D binary raster. However, when dealing with LiDAR datasets, we have to change that raster approach to a point-based data structure. In this scenario, many positions of a LiDAR matrix can be empty, that is, there are no points in those positions.

The $k^3$-lidar recursively divides the matrix into several equal-sized submatrices, by following the same strategy used by the $k^3$-tree. Again, this recursive subdivision is represented as a tree, where each node corresponds to a submatrix. As in the original $k^3$-tree, the division of the submatrices stops when the submatrix is empty (full of 0-bits, in the case of the $k^3$-tree) and when the process reaches the cells of the original matrix (submatrices of sixe $1\times 1 \times 1$), placed at level $\lceil log_k n \rceil$. In addition to these cases, the subdivision of the $k^3$-lidar also stops when the number of points in the processed submatrix is less than or equal to a given threshold $l$.  %That is,  a leaf node might contain a list of LiDAR points. 

Regarding the attributes of each point,  their values are compactly stored in our structure and can be efficiently retrieved when obtaining a point. Our structure stores the attributes defined in Point Data Record Format 0 (LAS Specification 1.4~\cite{las_14r14}) but it can be easily adapted to allow attributes defined in other formats.

\subsection{Data structures}\label{sec:k3-lidar:data_structures}

We use several data structures to represent the conceptual tree previously described:
\begin{itemize}
    \item \textbf{Tree structures ($T$ and $H$):} We use two data structures to represent the topology of the tree. $T$ is a bitmap, similar to that of the $k^3$-tree, but a 0 means that the submatrix is empty or the number of points does not exceed the threshold $l$. The $L$ bitmap of $k^3$-tree is not used. This is due to: i) in many cases, the $k^3$-lidar does not reach the last level of subdivision (level $\lceil log_k n \rceil$), as the division frequently stops before that level due to empty submatrices or submatrices containing $l$ points or less; and ii) in case of reaching the last level of the subdivision, leaf points require a more complex data structure than just a bitmap. 
           
        In addition to T, we use another bitmap, $H$, which has  one bit for each 0-bit in $T$, plus, if the last level is reached, as many bits as cells in that last level. %bits would have the $L$ bitmap of the $k^3$-tree. %one bit for  all position (we consider the last level as virtual 0s), 
    Each bit differentiates empty submatrices with those that contain some points. That is, for each $i$ such that  $T[i] \leftarrow 0$, we set $H[rank_0(T,i)] \leftarrow 0$ iff the submatrix is empty or $H[rank_0(T,i)] \leftarrow 1$ in other case. In the case of positions corresponding to the last level $\lceil log_k n \rceil$, a 1-bit means that the corresponding cell has points. 
    
    \item \textbf{Number of points ($N$):} We use a bitmap $N$ to store the number of points that each leaf contains. When $H[rank_0(T,i)] = 1$, we store in $N$ the number of points, in unary. %For example, value 3 is stored as \textit{001}.
    \item \textbf{Arrays of coordinates ($X$, $Y$, and $Z$):} Coordinates of a point $\langle x, y, z \rangle$ are stored in arrays X, Y, Z respectively. Points in  level $\lceil log_k n \rceil$ are not stored, since their position can be calculated during the descent through the tree. These values are encoded as local coordinates with respect to the node to which they belong. This results in smaller values than the original ones, and thus they are represented with DACs \cite{DACs}. DACs provide efficient random access to any position and a good compression ratio with small values.
    \item \textbf{Attributes:} Attributes, as intensity, return number, etc.,  are stored in separated arrays. They follow the same order as the coordinate arrays. The sequence of values are encoded with DACs or with a bitmap when the attribute can be represented with just one bit per point. 
    \item \textbf{Scale factor and Offset}: We convert real coordinates into positive integer values. Therefore, we use one scale factor and one different offset for each dimension. The scale factor is a float value that allows us to transform float values into integer values. The offset is an integer value that allows us to translate points to a coordinate system that starts at $\langle 0, 0, 0 \rangle$. In addition to converting negative numbers into positives, the offset also allows us to obtain smaller numbers in the values of the coordinates. For example, if the minimum value in $X$ is $1000$, we can move all points 1000 positions to the left, that is, the coordinate $1000$ would be the $0$, the $1001$ would be the $1$, and so on. These parameters use the same strategy than LAS/LAZ format.
\end{itemize}

\begin{figure}
    \includegraphics[width=\textwidth]{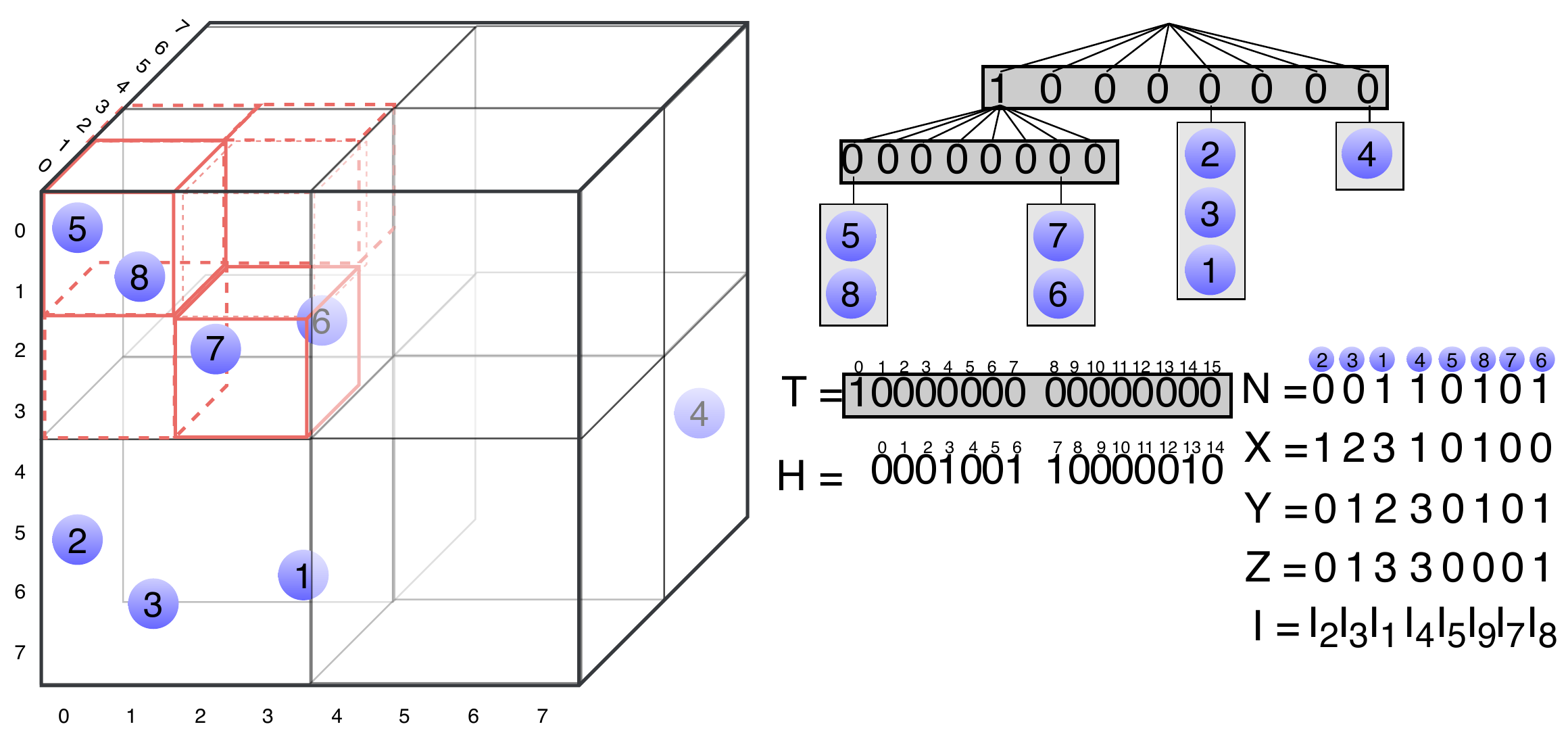}
    \caption{Example of a cloud of LiDAR points (left) where each point (circle) has an intensity $I_i$. Conceptual tree representation (top-right) and compact structures involved (bottom-right) in the construction of the $k^3$-lidar. This example uses $k=2$ and $l=3$ (maximum number of points at a leaf). For this example, only the intensity attribute is stored. The scale factor and the offset are not represent for clarity.} \label{fig:k3-lidar_lidar_example}
\end{figure}

\subsection{Construction}\label{sec:k3-lidar:construction}

Figure \ref{fig:k3-lidar_lidar_example} shows the recursive division of a cloud of LiDAR points (left), and the $k^3$-lidar representation (right-top). This examples uses $k=2$ and the maximum number of points in a leaf ($l$) is 3. Circles represent LiDAR points and they are identified with a unique identifier. In this case, each point only contains an intensity value labeled $I_{id}$, where $id$ is the identifier of that point. The algorithm starts by dividing the cube in $k^3 = 2^3 = 8$ submatrices of equal size. We add a child node of the root for each submatrix. The first child (bottom-left) has 4 points (\textit{5}, \textit{8}, \textit{7} and \textit{6}) and, since $4 > l$, we set $T[0] \leftarrow 1$. This node is then enqueued to be processed later. The second node is empty, thus $T[1] \leftarrow 0$ and $H[0] \leftarrow 0$. Note that the third submatrix (bottom-left) has 3 points (\textit{2}, \textit{3} and \textit{1}), thus, since $3 <=l$, we set $T[6] \leftarrow 0$,  $H[1] \leftarrow 1$, and add 3 in unary (001) to $N$. Moreover, local coordinates are stored into arrays $X$, $Y$, and $Z$ following a $z-$order. Point \textit{2} having global coordinates $\langle 5, 0, 0 \rangle$ becomes $\langle 1, 0, 0 \rangle$. Hence, $X[0] \leftarrow 1$, $Y[0] \leftarrow 0$, and $Z[0] \leftarrow 0$. This point has an intensity value of $I_2$, so we set $I[0] \leftarrow I_2$. We repeat the same process with nodes \textit{3} and \textit{1}, in that order. The algorithm continues with the rest of submatrices until reaching leaf nodes. In case of having more attributes, we would create a sequence $A$ for each attribute in an analogous way to the sequence of intensities $I$.

At the bottom-right of Figure \ref{fig:k3-lidar_lidar_example} we include the final structures that represent the $k^3$-lidar of the example.

\subsection{Query}\label{sec:k3-lidar:query}

In this section, we describe two queries designed for the $k^3$-lidar, which are of interest for LiDAR point clouds.

\subsubsection{Obtaining all points of a region (\textit{getRegion}):}

The $k^3$-lidar is able to obtain all points of a given region $\langle x_i, y_i, z_i \rangle$ $\times$ $\langle x_e, y_e, z_e \rangle$ by performing a top-down traversal of the tree from the root node. We follow the branches corresponding to submatrices that overlap with the region of interest. When a leaf is found, the coordinates of its points are checked and the attributes are only retrieved if the point is within the region. Due to the fact that points follow a $z$-order, some mechanisms are included to decrease the number of points checked. For instance, when we reach a point $\langle x', y', z' \rangle$ and $x' > x_e$, the process stops as we can assure that there are no more valid points in that node. 

Algorithm \ref{alg:k3-lidar:get_region} shows a pseudocode of the algorithm to solve this query. Let  $T$, $H$, $N$, $X$, $Y$, $Z$, and $1s\_in\_T$ be global parameters and $n\times n\times n$ the size of the dataset. Parameter $1s\_in\_T$ stores the total number of 1s in bitmap $T$ and was calculated previously. The parameter $result$ is the list of points returned by the query. Given a region $\langle x_i, y_i, z_i \rangle \times \langle x_e, y_e, z_e \rangle$, the first call of the algorithm is \textbf{getRegion}$(n, x_i, y_i, z_i,x_e, y_e, z_e, 0)$. 

Lines 1--3 run through each child node that overlaps the defined region.  $c\_pos$ (Line 4) is the position in $T$ of the current child node. The condition in Line 5 determines if the node is at  level $\lceil log_k n \rceil$ (which corresponds to submatrices of size $1 \times 1 \times 1$) or not.

Line 6 checks if the node is an internal node or a leaf node. In the first case, a recursive call is invoke (Lines 7--10). Function \textbf{getLocalCoordinates} converts the given region into local coordinates with respect to the current node. The position of its children is calculated as $rank(T, c\_pos) \cdot k^3$.

Lines 11--22 are executed when the algorithm reaches a leaf node. Line 12 counts the number of 0-bits in $T$ until position $c\_pos$, i.e the number of leaves until that position ($\#leaves$). The condition in Line 13 checks if the node is not empty. Line 14 counts the number of leaf nodes containing points ($\#ones$) until position $\#leaves$. 

Lines 15--17 obtain the positions of the first point and the last point of the current child node in arrays $X$, $Y$, and $Z$. Since the number of points has been inserted in unary code, each node corresponds to a 1-bit in the bitmap. With the \textit{select} operation, we obtain the position of the 1-bit corresponding to the previous node and the 1-bit of the current node. Intermediate positions are points of the current node. Finally, in Lines 18--22, the algorithm gets the coordinates of each point and checks if it belongs to the region of interest. If affirmative, the corresponding attributes are added and the point is inserted into the final result.

When the algorithm reaches a node in the level $\lceil log_k n \rceil$, Lines 24--32 are executed. Line 24 calculates the position in $H$, recall that level $\lceil log_k n \rceil$ is not represented in $T$. Lines 26--29 are equal to lines 14--17. Finally, for each point in the node, the algorithm retrieves its information and adds the point to the list. Observe that it is not necessary to obtain the local coordinates of the vectors $X$, $Y$ and $Z$.

\begin{algorithm}[t]
% \SetAlgoNoLine
\scriptsize
   \caption{{\bf getRegion}$(n,x_i,y_i,z_i,x_e,y_e,z_e,children\_pos)$ returns all LIDAR points from region $\langle x_i, y_i, z_i \rangle$ $\times$ $\langle x_e, y_e, z_e \rangle$}\label{alg:k3-lidar:get_region}
   
   \For{$x'\leftarrow\lfloor x_i / (n/k) \rfloor \ldots \lfloor x_e / (n/k) \rfloor$}{
       \For{$y'\leftarrow\lfloor y_i / (n/k) \rfloor \ldots \lfloor y_e / (n/k) \rfloor$}{
    
           \For{$z'\leftarrow\lfloor z_i / (n/k) \rfloor \ldots \lfloor z_e / (n/k) \rfloor$}{

                $c\_pos \leftarrow children\_pos + k\cdot k\cdot  x' + k\cdot y' + z'$ \\
                
                \eIf(\tcc*[h]{not at level $\lceil log_k n \rceil$ }){$(n/k) \neq 1$}
                {
        
                   \eIf(\tcc*[h]{internal node}){$T[c\_pos]=1$}{
                        
                        $\langle x_i', x_e' \rangle \leftarrow getLocalCoordinates(x', x_i, x_e, (n/k))$\\
                        $\langle y_i', y_e' \rangle \leftarrow getLocalCoordinates(y', y_i, y_e, (n/k))$\\
                        $\langle z_i', z_e' \rangle \leftarrow getLocalCoordinates(z', z_i, z_e, (n/k))$\\
                   
                         {{\bf getRegion}$(n/k,x_i',y_i',z_i',x_e',y_e',z_e', rank(T,c\_pos) \cdot k^3)$}
                    }
                    (~\tcc*[h]{leaf node})
                    {
                        $\#leaves \leftarrow rank_0(T,c\_pos)$\\
                        \If(\tcc*[h]{Node with points}){$H[\#leaves]=1$}{
                             $\#ones \leftarrow rank(H,\#leaves)$\\
                             \lIf{$\#ones=0$}{
                    		{$p\_init \leftarrow 0$}
                            }
                            \lElse{$p\_init \leftarrow select(N, \#ones)+1$}
                             {$p\_end \leftarrow select(N, \#ones+1)+1$}\\
                            
                            \For{$p'\leftarrow p\_init \ldots p\_end$}{
                                {$\langle p_x, p_y, p_z \rangle \leftarrow \langle X[p], Y[p], Z[p] \rangle$}\\
                                
                                 \If(){$\langle p_x, p_y, p_z \rangle$ within $\langle x_i, y_i, z_i \rangle$ $\times$ $\langle x_e, y_e, z_e \rangle$}{
                                    {$point \leftarrow retrieve\_attributes(p, \langle p_x, p_y, p_z \rangle)$}\\
                                    {ADD $point$ to $result$}\\
                                 }
                            } % END FOR p'
                        } % END IF check node
                    } % END else leaf node
                }
                 (~\tcc*[h]{last level})
                {
                    $\#leaves \leftarrow c\_pos - 1s\_in\_T$\\
                    \If(\tcc*[h]{Node with points}){$H[\#leaves]=1$}{
                        $\#ones \leftarrow rank(H,\#leaves)$\\
                             \lIf{$\#ones=0$}{
                    		{$p\_init \leftarrow 0$}
                            }
                            \lElse{$p\_init \leftarrow select(N, \#ones)+1$}
                             {$p\_end \leftarrow select(N, \#ones+1)+1$}\\
                            
                            \For{$p'\leftarrow p\_init \ldots p\_end$}{
       
                                {$point \leftarrow retrieve\_attributes(p, \langle x', y', z' \rangle)$}\\
                                {ADD $point$ to $result$}\\
                            } % END FOR p'
                        } % END IF check node
                }% END else last level
           } % END FOR Z
       } % END FOR Y
   } % END FOR X
\end{algorithm}

\subsubsection{Obtaining all points of a region filtered by attribute value (\textit{filterAttRegion}):} Given region $\langle x_i, y_i, z_i \rangle$ $\times$ $\langle x_e, y_e, z_e \rangle$ and a range of values for an attribute $[A_i, A_e]$, this query obtains all points within the defined region with values for the attribute between $A_i$ and $A_e$. Again, this query performs a top-down traversal of the tree. Unlike the query \textbf{getRegion}, when the algorithm reaches a leaf node, in addition to the coordinates, it also retrieves the attribute value of the point. 

%%%%%%%%%%%%%%%%%%%%%%%%%%%%%%%%%%%%%%%%%%%%%%%%%%%%%%%%%%%%%%%%%%%%%%
\section{Experimental evaluation}\label{sec:experimental}
%%%%%%%%%%%%%%%%%%%%%%%%%%%%%%%%%%%%%%%%%%%%%%%%%%%%%%%%%%%%%%%%%%%%%%
We ran some experiments as a proof of concept of the good properties of our proposed data format. More concretely, we compare the space and time results obtained by $k^3$-lidar, to those obtained when using LAS/LAZ formats. 
%More concretely, we measured space requirements and execution time for answering the. % In this section, we measure the space and the result obtained by  in comparison with the LAS/LAZ format.

% \subsection{Experimental framework}\label{sec:experimental:framework}

Table \ref{tab:experimental:datasets} shows the description of the LiDAR point clouds used in the experimental evaluation. We use five different datasets coming from two different sources, an airborne LiDAR and a mobile laser scanning:
\begin{itemize}
    \item Three datasets were created from the union of different files of the \textit{Plan Nacional de Observaci{\'o}n del Territorio}\footnote{http://pnoa.ign.es/productos\_lidar} (PNOA). Each tile (file) represents an area of Spanish territory of size 2$\times$2 km with a minimum density of 0.5 points/m$^2$. \texttt{PNOA-small} is composed of 4 tiles and represents an area of 16 km$^2$, \texttt{PNOA-medium} is composed of 9 tiles and represents an area of 36 km$^2$, and \texttt{PNOA-large} is composed of 23 tiles and represents an area of 92 km$^2$. The number of points can vary from one tile to another. These datasets contain the following attributes: intensity (with values between 0 and 255), return number (with values between 1 and 4), number of returns (with values between 1 and 4), edge of flight line (with values between 0 and 1), scan direction flag (with values between 0 and 1), classification (with values between 1 and 7), scan angle rank (with values between -24 and 28), and point source ID (with values between 175 and 227). 
    
    \item We use datasets {\tt TUB1} and {\tt FireBrigade} from the ISPRS benchmark on indoor modelling\footnote{\url{http://www2.isprs.org/commissions/comm4/wg5/benchmark-on-indoor-modelling.html}} \cite{khoshelham2017isprs}. {\tt TUB1} point cloud was captured in one of the buildings of the Technische Universität Braunschweig, Germany, and {\tt FireBrigade} was captured in the office of fire brigade in Delft, The Netherlands. These datasets do not contain values for any additional attribute, just the point coordinates. The level of clutter, defined as the amount of points belonging to elements that do not constitute the building structures, is low for {\tt TUB1} and high for {\tt FireBrigade}.
\end{itemize}

In all cases, LiDAR points were converted to Point Data Record Format 0 (LAS Specification 1.4). Then we created indexes for the LAZ files using the \textit{lasindex} tool of LAStools\footnote{\url{https://github.com/LAStools/LAStools}}. These indexes are able to index the $x$ and $y$ dimensions to improve the query time. LASLib library\footnote{\url{https://github.com/LAStools/LAStools/tree/master/LASlib}} was used to execute queries on LAZ files. The $k^3$-lidar was configured with $k=2$ and $l=100$.

All the experiments were run on an isolated Intel\textsuperscript{\textregistered} Core\textsuperscript{TM} i7-3820 CPU @ 3.60 GHz (4 cores) with 10 MB of cache, and 64 GB of RAM. It ran Debian 9.8 \textit{Stretch}, using gcc version 6.3.0 with -03 option. 

\begin{table}[t]
  \centering
  \caption{Datasets description. We show the number of points, the minimum and maximum values for the real coordinates $x,y,z$, and the maximum $X,Y,Z$ values, after the coordinates are converted (scaled and translated using an offset).}\label{tab:experimental:datasets}
    \begin{tabular}{|l|r|r|r|r|r|}
\cline{2-6}     
\multicolumn{1}{r|}{}& \texttt{PNOA-small} & \texttt{PNOA-medium} & \texttt{PNOA-large} & \texttt{TUB1} & \texttt{FireBridge} \\\hline
	\#points & 13,265,144 & 25,108,130 & 52,627,503 & 32,597,694 & 10,406,389 \\\hline\hline
    min \textit{x} (real) & 546,000.00 &  544,000.00 &  542,000.00 & -9.90   & 44.32  \\
    \hline  
    max \textit{x} (real) &  549,999.99 & 549,999.99 &  551,999.99 & 5.76  &  58.40 \\
    \hline 
    min \textit{y} (real) & 4,798,000.00 & 4,798,000.00  & 4,794,000.00  & -24.52 & 23.10  \\
    \hline
    max \textit{y} (real) &  4,801,999.99  & 4,803,999.99 & 4,805,020.45 & 18.30  & 77.67 \\
    \hline 
	min \textit{z} (real) & -39.14 & -162.43 & -162.43 & -1.46   & 1.27  \\
	\hline
	max \textit{z} (real) &  179.58 &  1005.05 & 1005.05 &  1.10   & 12.04 \\
    \hline\hline
    max \textit{X} (converted)    & 3,999,990     & 5,999,990      & 9,999,990      & 1,000,000,000     & 1,000,000,000 \\
    \hline
    max \textit{Y} (converted)     & 3,999,990     & 5,999,990      & 11,020,450     & 1,000,000,000      & 1,000,000,000 \\
    \hline
    max \textit{Z} (converted)      & 218,720      & 1,167,480      & 1,167,480     &1,000,000,000     & 1,000,000,000 \\\hline
        \end{tabular}%
  \label{tab:datasetsdesc}%
\end{table}%

% \subsection{Comparison with related work}\label{sec:experimental:comparison}

The comparison of space is shown in the first columns of Table \ref{tab:experimental:results}. LAZ files obtain the best results in the three cases, around 65\% less than $k^3$-lidar, which in turn, needs around 53\%  less space than the uncompressed LAS. 

%\begin{table}
%\centering
%\caption{Comparison of space. }\label{tab:experimental:space}
%    \begin{tabular}{|l|r|r|r|}
%        \hline
%        \textbf{Dataset} & \textbf{LAS} & \textbf{LAZ} & \textbf{$k^3$-lidar}\\
%        \hline
%        Small & 254 MB & \textbf{43 MB} & 119 MB\\
%        Medium & 479 MB & \textbf{80 MB} & 225 MB\\
%        Large & 1004 MB & \textbf{173 MB} & 471 MB\\
%        \hline
%    \end{tabular}
%\end{table}

% k3-lidar configuration: k=2, l=100
% Number of queries
% Small ->  208/500      196/200
% Medium -> 135/500      108/200
% Large -> 165/500       105/100

Table \ref{tab:experimental:results} also shows the query times. We generated 500 random regions of different sizes. However, the LAZ software failed in many queries (the reported result contains 0 points). Therefore we only considered the times of those techniques that worked properly, that is, LAZ and $k^3$-lidar. Our proposal is around 5 times faster than LAZ in {\em GetRegion} queries for {\tt PNOA} datasets and from 16--23 times faster for {\tt ISPRS} datasets. {\em FilterAttRegion} queries were only executed over {\tt PNOA} datasets filtering by the intensity attribute, as {\tt TUB1} and {\tt FireBrigade} only contain the coordinates of the points, but do not include any other attribute values. For  {\tt PNOA} datasets, our proposal outperforms LAZ format, as queries are solved 5--10 times faster.

\begin{table}
\centering
\caption{Comparison of the space (MB) and average time (in milliseconds) for queries \textit{getRegion} and \textit{FilterAttRegion}. }\label{tab:experimental:results}
    \begin{tabular}{|l||l||r|r|r||r|r||r|r|}
        \hline
         \multirow{ 2}{*}{Dataset} &  \multirow{ 2}{*}{\# points} & \multicolumn{3}{|c||}{\textbf{Space (MB)}} &
         \multicolumn{2}{|c||}{\textbf{GetRegion (ms)}} & \multicolumn{2}{|c|}{\textbf{FilterAttRegion (ms)}}\\
        \cline{3-9}
         &  & \textit{LAS} & \textit{LAZ} & \textit{$k^3$-lidar} & \textit{LAZ} & \textit{$k^3$-lidar} & \textit{LAZ} & \textit{$k^3$-lidar}\\
        \hline
        \hline
        \texttt{PNOA-small} & 13,265,144  & 254  & \textbf{43} & 119 & 1,524 & \textbf{249} & 1,517 & \textbf{145}\\
        \texttt{PNOA-medium} & 25,108,130  & 479  & \textbf{80} & 225 & 2,521 & \textbf{424} & 2,655 & \textbf{374}\\
        \texttt{PNOA-large} & 52,627,503  & 1004  & \textbf{173} & 471 &6,859 & \textbf{1,189} & 6,283 & \textbf{1,264}\\
        \texttt{TUB1} & 32,597,694  & 622  & \textbf{196} & 304 & 6,145 & \textbf{383} & -- & -- \\
        \texttt{FireBrigade} & 10,406,389  & 199  & \textbf{77} & 100 & 1,717 & \textbf{74} & -- & -- \\
        \hline
    \end{tabular}
\end{table}

\section{Conclusions}\label{sec:conclusions}

In this work, we address the main drawback of the LAZ format for LiDAR data, which is its high executing times when answering to queries that retrieve a subset of points using constraints over the third dimension. LAZ is penalized by the fact that  decompression is performed by blocks and  the index only covers the X and Y coordinates. 

We propose a new representation for LiDAR point clouds, denoted $k^3$-lidar, which is able to decompress random points of the cloud and, as it is based on a compact version of an octree, the $k^3$-tree, it can index the three dimensions. This implies significant improvements in the querying times, ranging from 5 times to more than one order of magnitude faster.

The future work will cover two main lines. First, regarding the space requirements, $k^3$-lidar compresses 65\% less than LAZ. Therefore, the use of previous ideas from the original $k^2$-tree and $k^3$-tree, such as using different $k$ values in different levels, or further compressing frequent submatrices, will be studied.
The other line is the indexation of more dimensions, to include the attribute values stored at the points. The $k^2$-raster is a compact data structure designed for raster data that not only indexes the data spatially, but it also indexes the values stored at the cells of the raster. Our aim is to apply similar ideas to address the indexation of LiDAR data.

\bibliographystyle{splncs04}
\bibliography{main}

\section*{Appendix}

\begin{figure}[t]
    \centering
    \includegraphics[width=0.5\textwidth]{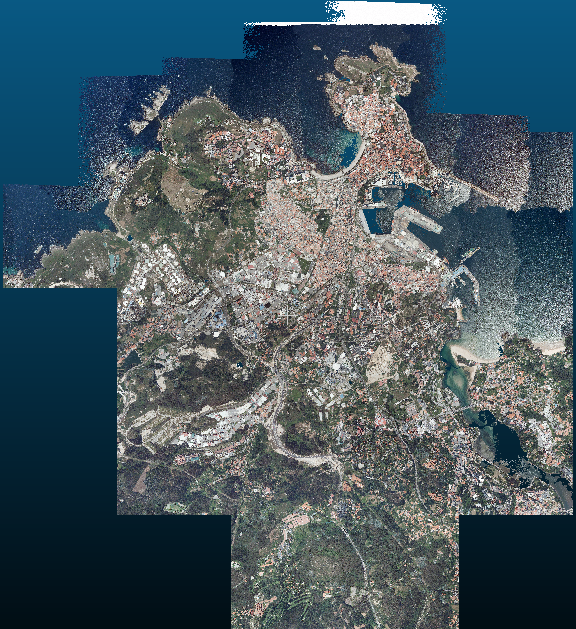}
    \caption{Visualization of the dataset labeled as \textit{Large}.}   \label{fig:experimental:large_dataset}
\end{figure}

\begin{figure}
    \centering
    \begin{subfigure}[b]{0.6\textwidth}
    \includegraphics[width=\textwidth]{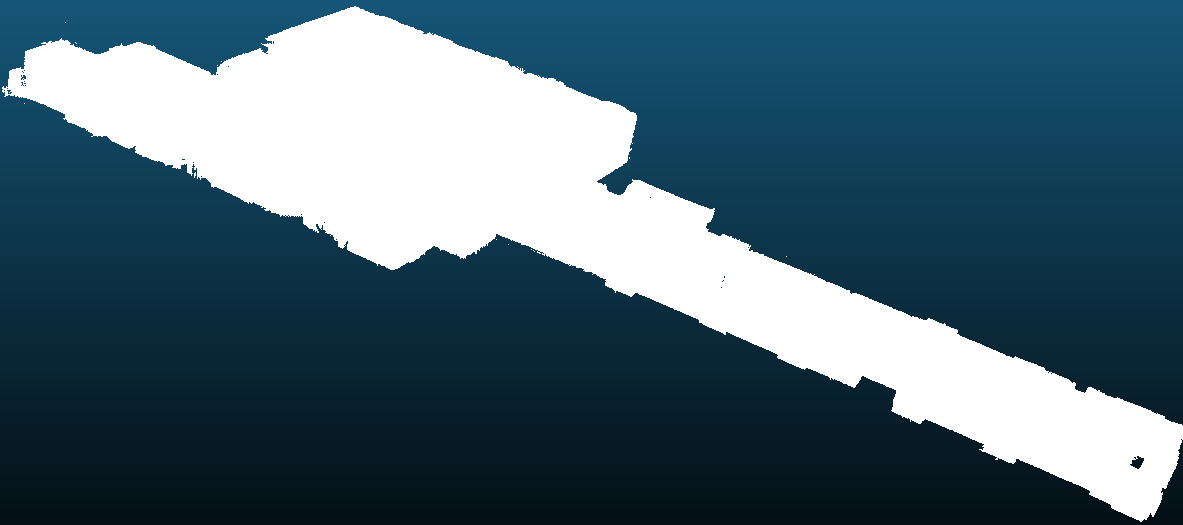}
    \caption{{\tt TUB1} point cloud visualization}
    \end{subfigure}    
    \begin{subfigure}[b]{0.6\textwidth}
    \includegraphics[width=\textwidth]{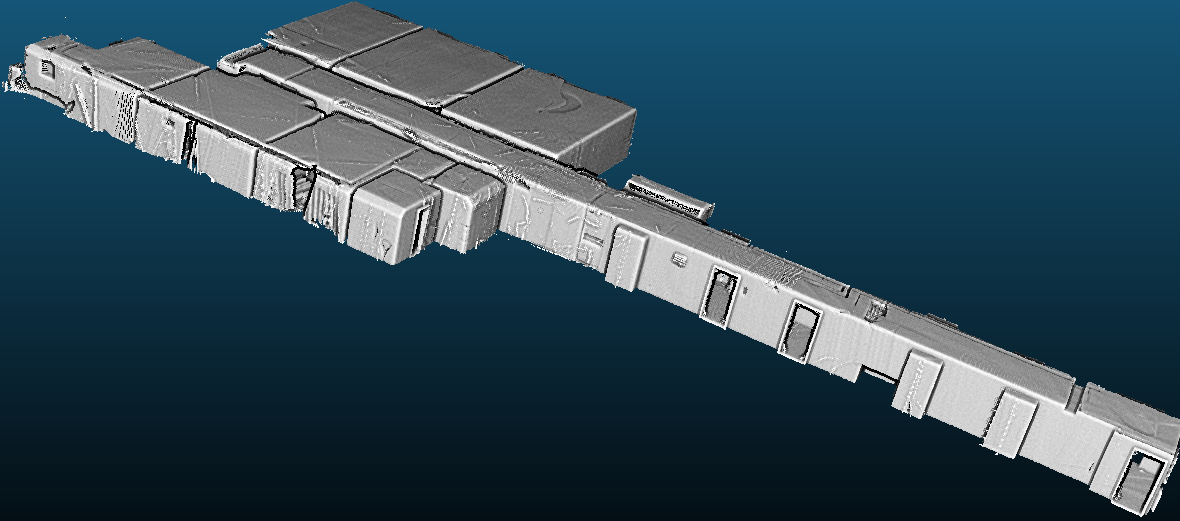}
    \caption{{\tt TUB1} eye-dome lighting visualization}
    \end{subfigure}
    \begin{subfigure}[b]{0.6\textwidth}
    \includegraphics[width=\textwidth]{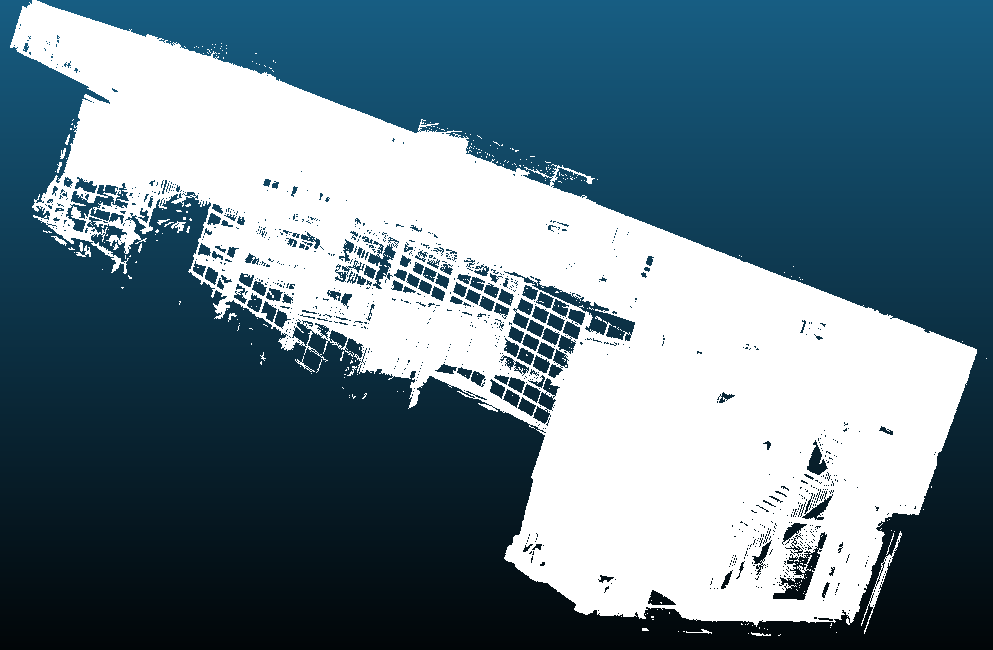}
    \caption{{\tt FireBrigade} point cloud visualization}
    \end{subfigure}
    \begin{subfigure}[b]{0.6\textwidth}
    \includegraphics[width=\textwidth]{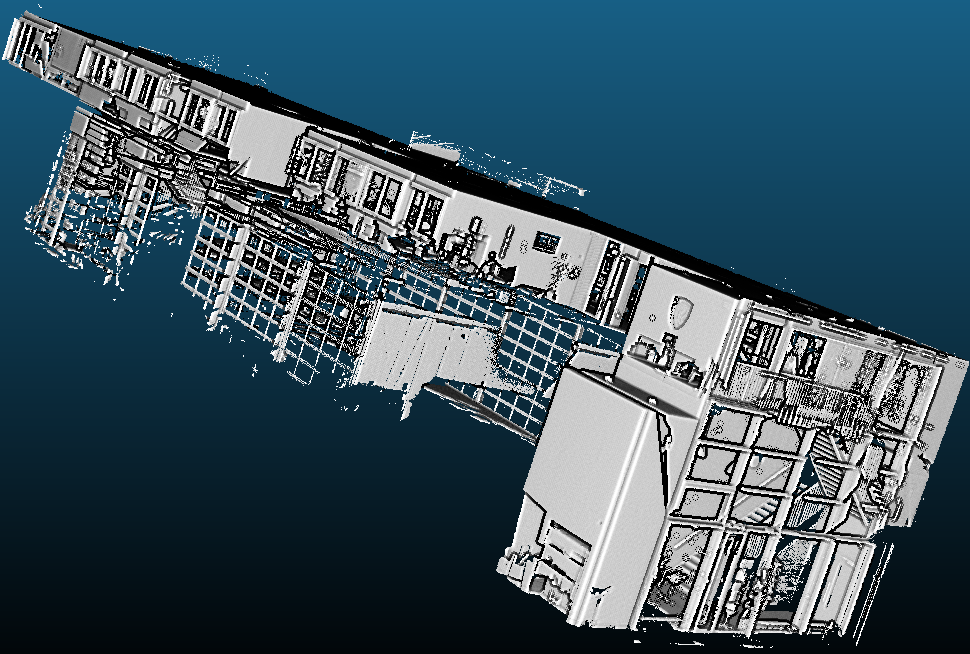}
    \caption{{\tt FireBrigade} eye-dome lighting visualization}
    \end{subfigure}
    \caption{Visualization of datasets {\tt TUB1} and {\tt FireBrigade}. We include the point cloud visualization and also an eye-dome lighting (EDL) visualization. EDL is a non-photorealistic, image-based shading technique designed to improve depth perception in scientific visualization images \cite{EDL}.}   \label{fig:experimental:tub}
\end{figure}

To better understand the nature of the datasets, we show a visualization of \texttt{PNOA-large} in Figure \ref{fig:experimental:large_dataset}, and visualizations of the point clouds {\tt TUB1} and {\tt FireBrigade} in Figure \ref{fig:experimental:tub}. 
 
\end{document}